\documentclass[runningheads]{svmult}

\usepackage{makeidx}   
\usepackage{graphicx}  
\usepackage{subeqnar}  
\usepackage{multicol}  
\usepackage{physprbb}  
\makeindex             

\begin{document}
\title*{From Darkness to Light: The First Stars in the Universe}
\toctitle{From Darkness to Light: The First Stars in the Universe}
\titlerunning{From Darkness to Light: The First Stars in the Universe}
\author{Volker Bromm\inst{1}
\and Paolo S. Coppi \inst{2}
\and Richard B. Larson\inst{2}}
\authorrunning{Volker Bromm et al.}
%
%
\institute{Harvard-Smithsonian Center for Astrophysics\\ Cambridge, MA 02138, USA
\and Astronomy Department, Yale University\\
     New Haven, CT 06520-8101, USA}      

\maketitle              

\begin{abstract}
Paramount among the processes that ended the cosmic `dark ages'
must have been the formation of the first generation of stars. In order to
constrain its nature, we investigate the
collapse and fragmentation of metal-free gas clouds. We
explore the physics of primordial star formation by means of three-dimensional
simulations of the dark matter and gas components, using smoothed particle
hydrodynamics.
We find characteristic values for the
temperature, $T\sim$ a few 100 K, and the density, $n\sim 10^{3}-10^{4}$ cm$^{-3}$,
characterising the gas at the end of the initial free-fall phase.
The corresponding Jeans mass is $M_{J}\sim
10^{3}M_{\odot}$. The existence of these characteristic values has a
robust explanation in the microphysics of H$_{2}$ cooling, and is not very
sensitive to the cosmological initial conditions.
These results suggest that the first stars might have been quite massive,
possibly even very massive with $M_{\ast}>100 M_{\odot}$.
\end{abstract}

\section{Introduction}

The history of the universe proceeded from an extremely uniform initial
state to the highly structured present-day one. When did the crucial
transition from simplicity to complexity first occur? Recently, this
question has become the focus of an intense theoretical
effort (e.g.,~\cite{abn,bl,bcl1,bcl2,bfcl,bkl,lb,nu}). 
There must have been a time, between the last scattering
of the CMB photons at $z\sim 1000$ and the formation of the first luminous
objects at $z > 6$, when the universe contained no visible light. This era
has been called the cosmic `dark ages', and the key question is how and
when it ended (e.g.,~\cite{rees}).

In the context of hierarchical scenarios of structure
formation, as specified by a variant of the cold dark matter (CDM) model,
the collapse of the first baryonic objects is expected at redshifts
$ z\simeq 50-10$, involving dark matter halos of mass $\sim 10^{6}M_{\odot}$
\cite{teg}.
The question arises how one can make any progress in understanding 
primordial star formation, given the lack of direct observational
constraints. The physics of the first stars, however, is characterized 
by some important simplifications, as compared to the extreme complexity
of present-day star formation~\cite{lar}. The absence of metals,
and consequently of dust, leaves atomic and molecular hydrogen as the main
agent of radiative cooling and the source of opacity. Magnetic fields were
likely to be dynamically insignificant, prior to the onset of efficient
(stellar) dynamo amplification. The chemistry and heating
of the primordial gas was not yet complicated by the presence of a UV
radiation background. The intergalactic medium (IGM)
must have been a rather quiescent place,
with no source to sustain turbulent motion, as long as the
density perturbations remained in their linear stage. Only after the
explosion of the first supernovae, and the associated input of mechanical
and thermal energy, is this state of primordial tranquility bound to change
\cite{hl}. Therefore, the
physics of primordial star formation is mainly governed by gravity, thermal
pressure, and angular momentum. This situation renders the problem
theoretically more straightforward and tractable than the highly complex
present-day case which continues to defy attempts to formulate a 
fundamental theory of star formation.
Finally, the initial conditions for the collapse of a primordial star 
forming cloud are given by the adopted model of cosmological structure
formation. 

The importance of the first stars and quasars derives from the crucial
feedback they exert on the IGM. A generation of
stars which formed out of primordial, pure H/He gas (the so-called
Population III) must have existed, since heavy elements can only be
synthesized in the interior of stars. Population III stars, then, were
responsible for the initial enrichment of the IGM with heavy elements. From
the absence of Gunn-Peterson absorption in the spectra of high-redshift
quasars, we know that the universe has undergone a reionization event
at $z > 6$. UV photons from the first stars, perhaps
together with an early population of quasars, are expected to have contributed
to the reionization of the IGM \cite{hl,fer,meshr}.

To probe the time
when star formation first started entails observing at redshifts $z > 10$.
This is one of the main purposes of the {\it Next Generation
Space Telescope} (NGST) which is designed to reach $\sim$ nJy sensitivity
at near-infrared wavelengths. In preparation for this upcoming
observational revolution, the study of the first stars is very timely,
providing a theoretical framework for the interpretation of what NGST
might discover, less than a decade from now.

\section{Simulations}
Our code is based on a version
of TREESPH which combines
the Smoothed Particle Hydrodynamics (SPH) method
with a hierarchical (tree) gravity solver.
To study primordial gas, we have made a number of additions.
Most importantly, radiative cooling due to hydrogen molecules
has been taken into account. In the absence of metals, H$_{2}$ is the
main coolant below $\sim 10^{4}$ K, the typical
temperature range in collapsing Population III objects. 
The efficiency of H$_{2}$ cooling is very sensitive to the H$_{2}$ abundance.
Therefore, it is necessary to compute the nonequilibrium evolution
of the primordial chemistry
(see~\cite{bcl2}
for details).

We have devised an algorithm to merge SPH particles
in high density regions to overcome the otherwise prohibitive
timestep limitation, as enforced
by the Courant stability criterion.
To follow the simulation for a few dynamical times, we 
allow SPH particles to merge into more
massive ones, provided they exceed a pre-determined
density threshold, typically $10^{8}-10^{10}$ cm$^{-3}$~\cite{bcl2}.

We have carried out a comprehensive survey of the relevant parameter
space, and focus in the following on one select example that is 
representative for the overall results.

\subsection{The Fragmentation of Primordial Gas}

We model the site of primordial star formation as an isolated overdensity,
corresponding to a high$-\sigma$ peak in the random field
of density perturbations.
The numerical simulations are initialized at $z_{i}=100$ by endowing a
spherical region, containing dark matter and gas, with a uniform density
and Hubble expansion. Small-scale density fluctuations are imprinted
on the dark matter according
to the CDM power spectrum.
We assume that the halo is initially in
rigid rotation with a given angular velocity, 
prescribed in
accordance with the prediction for the spin parameter $\lambda$,
as found in cosmological N-body simulations.

\begin{figure}[h]
\begin{center}
\includegraphics[width=0.8\textwidth]{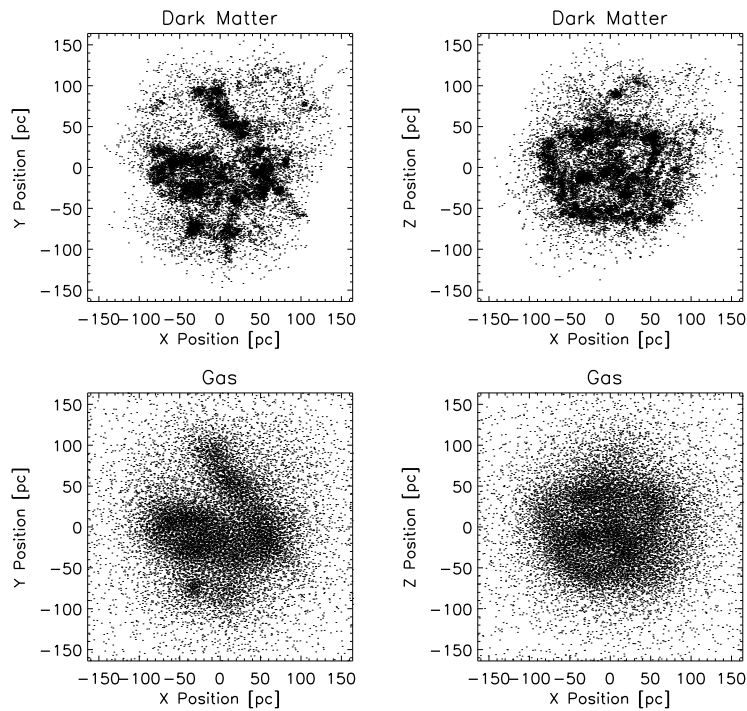}
\end{center}
\caption[]{Collapse of a primordial star forming cloud (from~\cite{bcl2}).
The halo has a total mass of $2\times 10^{6}M_{\odot}$, and will collapse
at $z_{vir}\simeq 30$. Shown here is the morphology at $z=33.5$.
{\it Top row:} DM particles. {\it Bottom row:} Gas particles.
{\it Left panels:} Face-on view. {\it Right panels:} Edge-on view.
The DM has developed significant substructure, and the baryons are just
beginning to fall into the corresponding potential wells.}
\label{eps1}
\end{figure}

Fig. 1 shows the situation
at $z=33.5$, briefly before the virialization of the dark matter. In 
response to the initially imprinted $k^{-3}$-noise, the dark matter has
developed a pronounced substructure. 
The baryons have just begun to fall into the potential wells which
are created by the DM substructure. Thus, the DM imparts a `gravitational
head-start' to certain regions of the gas, which subsequently act as
the seeds for the formation of high-density clumps.

At the end of the free-fall phase, the gas has developed
a very lumpy, filamentary structure in the center of the DM potential. The gas
distribution is very inhomogeneous, and the densest regions are gravitationally
unstable. The ensuing runaway collapse leads to the formation of
high-density sink particles or clumps. These clumps are formed with
initial masses close to $M\sim 10^{3}M_{\odot}$, and subsequently
gain in mass by the accretion of surrounding gas, and by merging with
other clumps (see Fig. 2).

\begin{figure}[h]
\begin{center}
\includegraphics[width=0.8\textwidth]{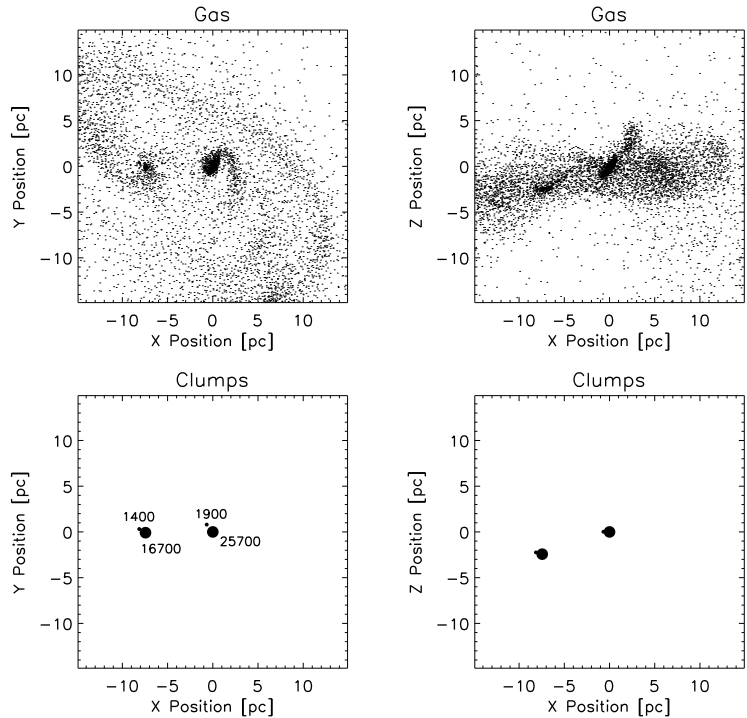}
\end{center}
\caption[]{Fragmentation into high density clumps (from~\cite{bcl2}).
Shown here is the situation at $z=28.9$ in the innermost 30 pc.
{\it Top row:} The remaining gas in the diffuse phase.
{\it Bottom row:} Distribution of clumps. The numbers next to the dots
denote clump mass in units of $M_{\odot}$.
{\it Left panels:} Face-on view. {\it Right panels:} Edge-on view.
The gas has settled into an irregular configuration with two
dominant clumps. The clumps are formed with initial masses of $\sim
10^{3}M_{\odot}$, and subsequently grow in mass by accretion and
merging with other clumps, up to $\sim 20,000 M_{\odot}$.}
\label{eps2}
\end{figure}

There is a good physical reason for the emergence of high-density clumps
with initial masses $\sim 10^{3}M_{\odot}$. To understand this, consider
the thermodynamic and chemical state of the gas, as 
summarized in Fig. 3. Since the abundances, temperature and density
are plotted for every SPH particle, this mode of presentation has an
additional dimension of information: Particles accumulate (`pile up') in those
regions of the diagram where the evolutionary timescale is long.
In panel (c) of Fig. 3, one can clearly discern such a preferred state
at temperatures of a few 100 K, and densities of $10^{3}-10^{4}$ cm$^{-3}$.
These characteristic values have a straightforward physical explanation
in the microphysics
of H$_{2}$ cooling. A temperature of $T\sim 100-200$ K is the minimum one
attainable via H$_{2}$ cooling. The corresponding critical density, beyond
which the H$_{2}$ rotational levels are populated according to LTE,
is then $n_{crit}\simeq 10^{3}-10^{4}$ cm$^{-3}$. 
Due to the now inefficient cooling, the gas `loiters'
and passes through a phase of quasi-hydrostatic, slow contraction.

\begin{figure}[h]
\begin{center}
\includegraphics[width=0.8\textwidth]{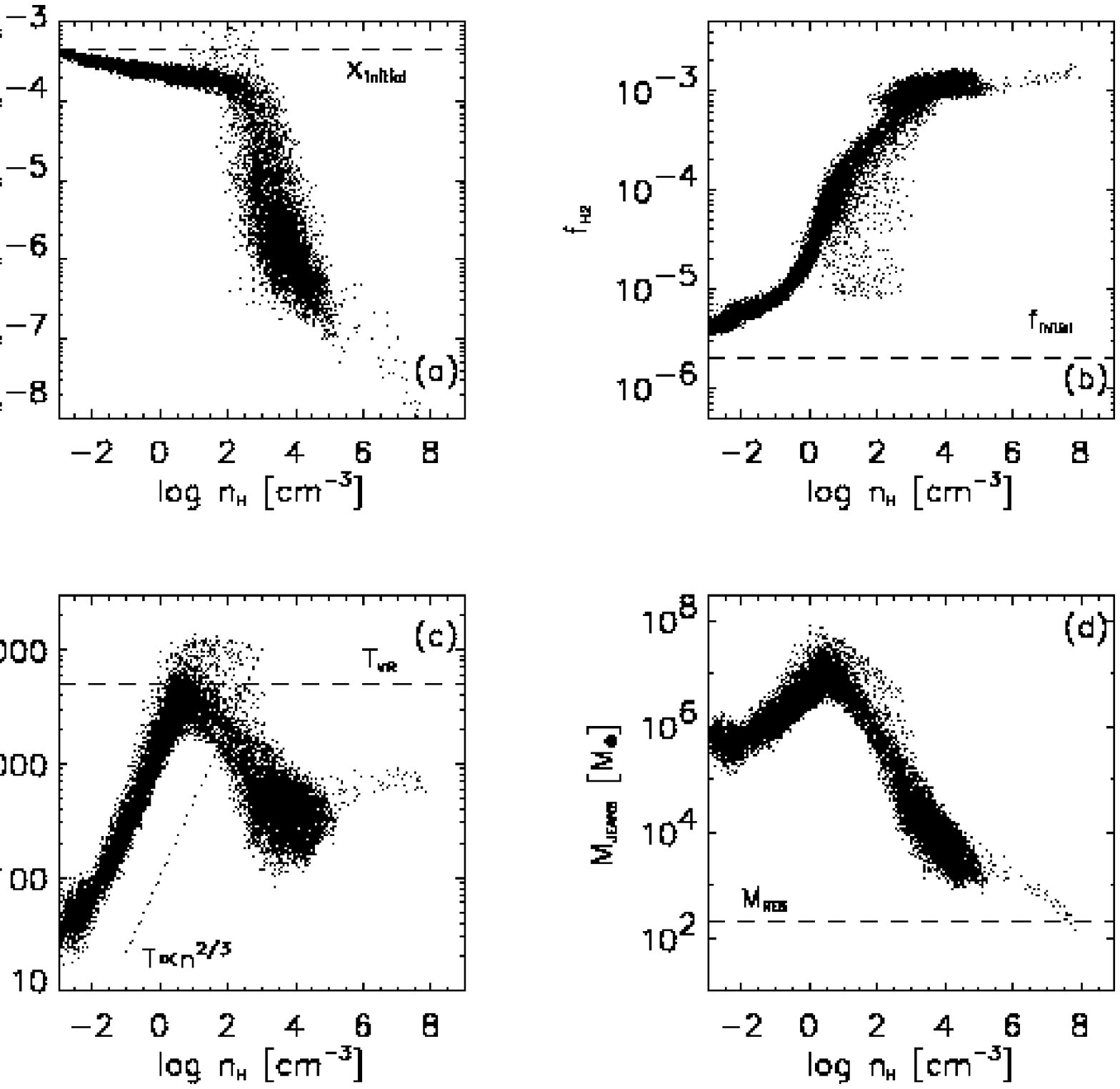}
\end{center}
\caption[]{Gas properties at $z=31.2$ (from~\cite{bcl2}).
{\bf (a)} Free electron abundance vs. hydrogen number density (in cm$^{-3}$).
{\bf (b)} Hydrogen molecule abundance vs. number density. 
{\bf (c)} Gas temperature vs. number density. At densities below $\sim 1$ cm$^
{-3}$, the gas temperature rises because of adiabatic compression until
it reaches the virial value of $T_{vir}\simeq 5000$ K.
At higher densities, cooling due to H$_{2}$
drives the temperature down again, until the gas settles into a quasi-
hydrostatic state at $T\sim 300$ K and $n\sim 10^{4}$ cm$^{-3}$.
Upon further compression due to the onset of the gravitational
instability, the temperature experiences a modest rise again.
{\bf (d)} Jeans mass (in $M_{\odot}$) vs. number density. The Jeans mass
reaches a value of $M_{J}\sim 10^{3}M_{\odot}$ for the quasi-hydrostatic
gas in the center of the DM potential well.}
\label{eps3}
\end{figure}

To move away from
this loitering regime, and to attain higher densities, the gas has to 
become gravitationally unstable. 
Evaluating
the Jeans mass for the characteristic values $T\sim 200$ K and
$n\sim 10^{3}-10^{4}$ cm$^{-3}$ results in $M_{J}\sim 10^{3}M_{\odot}$.
When enough gas has accumulated in a given region to satisfy $M > M_{J}$,
runaway collapse of that fluid region ensues. We find
that the gas becomes self-gravitating ($\rho_{\mbox{\scriptsize B}} >
\rho_{\mbox{\scriptsize DM}}$) coincident with the onset of the Jeans
instability. 

\subsection{Protostellar Collapse}
To further constrain the characteristic mass scale for Population III
stars, we have investigated the collapse of a clump to even higher
densities. As our starting configuration, we select the first region to
undergo runaway collapse in one of the lower resolution simulations
and employ a 
technique of refining the spatial resolution in the vicinity of this region,
which in the unrefined simulation would have given rise to the
creation of a sink particle.

\begin{figure}[h]
\begin{center}
\includegraphics[width=0.7\textwidth]{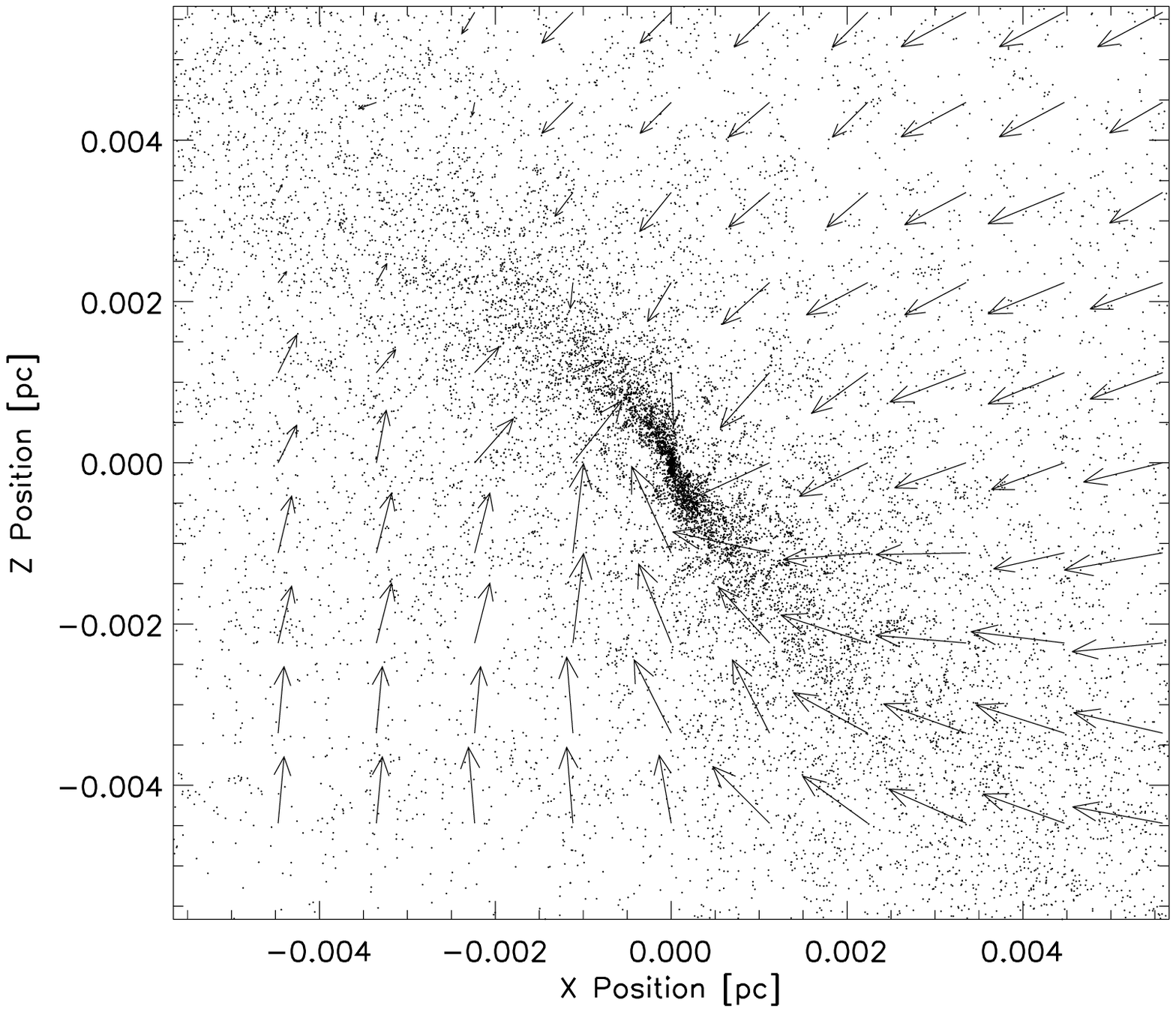}
\end{center}
\caption[]{
Gas morphology and kinematics in the vicinity of the density 
maximum.
Shown is the situation $\sim 5000$ yr after the onset of runaway collapse
in a box of linear
size $\sim 2500$ AU. The small dotted symbols give an indication of the gas
density, and the overplotted arrows depict the velocity field in the x-z plane.
The length of an arrow scales with speed such that the largest one corresponds
to $\sim 14.5$ km s$^{-1}$. It is evident that a highly concentrated,
spindle--like structure has formed in the center which comprises a few tens
of solar masses at this instant. The surrounding flow field is supersonic
with typical Mach numbers of $\sim 3-5$.
}
\label{eps5}
\end{figure}

In the refined simulation, 
three-body reactions become
important, and lead to the almost complete conversion of the gas into
molecular form~\cite{pss}.

Only $\sim 5000$ yr after the onset of runaway collapse,
the central, highest-density region has already evolved 
significantly. As can be seen in Fig. 4, an elongated, spindle-like
structure has formed, which comprises a mass of $\sim 20 M_{\odot}$, and
has a characteristic size of $L_{char}< 10^{-4}$ pc $\simeq 20$ AU.
By examining the surrounding velocity field, which is characterized by
Mach numbers of $\sim 3 - 5$, it is evident that matter continues to
fall onto the central object. To derive an estimate for the accretion rate,
we consider the average mass flux, $<\!\rho v_{r}\!>$, through a spherical surface
around the density maximum with radius $r=10^{-3}$ pc. Here, $v_{r}$ is the
radial velocity component, and velocities are measured relative to the 
density maximum. Assuming spherical accretion, one finds for the
accretion rate
\begin{equation}
\dot{M}_{acc}=-4\pi r^{2}\!<\!\rho v_{r}\!>\sim 1 M_{\odot}\mbox{\ yr$^{-1}$}
\mbox{\ \ \ .}
\end{equation}
This is likely to be an overestimate, since the assumption of spherical
symmetry is only a rough approximation to the complex kinematics of the flow.
Nevertheless, it is clear that the central object will rapidly grow in mass,
on a timescale $t_{acc}\sim M/\dot{M}_{acc}\sim 20$ yr.
We find no indication
for further subfragmentation in this simulation. 
These results suggest that the first stars might have been quite massive,
possibly even very massive with $M_{\ast}>100 M_{\odot}$.

Important caveats remain, however.
The question of how massive
the incipient star in the center of the collapsing clump will eventually be,
cannot be answered with any certainty at present. Our attempts in doing so
are foiled by our ignorance of the complex and rather unexplored physics
of accretion from a dust free envelope. 
This, then, is the frontier of our
current knowledge~\cite{op}.



%

\end{document}